\documentclass{article}
\usepackage[T1]{fontenc}
\usepackage[utf8]{inputenc}
\usepackage[]{ismir} 
\usepackage{amsmath,cite,url}
\usepackage{graphicx}
\usepackage{color}
\usepackage{enumitem}

\title{Modeling the Difficulty of Saxophone Music}

\oneauthor
  {Šimon Libřický, Jan Hajič jr.}
  {Institute of Formal and Applied Linguistics, Charles University \\ \texttt{librickysimon@gmail.com, hajicj@ufal.mff.cuni.cz}}

\def\authorname{Š. Libřický, and J. Hajič jr.}

\usepackage[bookmarks=false,pdfauthor={\authorname},pdfsubject={\pdfsubject},hidelinks]{hyperref}

\begin{document}

\maketitle

\begin{abstract}
In learning music, difficulty is an important factor in choice of repertoire,
choice of tempo, and structure of practice. These choices are typically done
with the guidance of a teacher; however, not all learners have access to one.
While piano and strings have had some attention devoted to automated difficulty estimation,
wind instruments have so far been under-served.
In this paper, we propose a method for estimating the difficulty of pieces for winds 
and implement it for the tenor saxophone.
We take the cost-of-traversal approach, modelling the part as a sequence of transitions -- note pairs. 
We estimate transition costs from newly collected recordings of trill speeds,
comparing representations of saxophone fingerings at various levels of expert input.
We then compute and visualise the cost of the optimal path through the part, at a given tempo.
While we present this model for the tenor saxophone, 
the same pipeline can be applied to other woodwinds, 
and our experiments show that with appropriate feature design, 
only a small proportion of possible trills is needed to estimate the costs well.
Thus, we present a practical way of diversifying the capabilities of MIR in music education
to the wind family of instruments.
\end{abstract}

\section{Introduction and system overview}
\label{sec:introduction}

One essential element of studying music is choosing appropriate repertoire. Out of the
factors that make a piece appropriate for a learner, its difficulty is among the most important:
too much of a challenge, or not enough of one, can seriously damage the learning process \cite{guadagnoli2004challenge, Bugos2014}.
A major factor governing the difficulty of executing an instrumental part is also its target tempo,
which must be determined.
In addition to estimating the difficulty of entire pieces, an overview of which parts of a piece are difficult (and hence will likely require the most practice)
might be useful for any learner. 
While teachers are of course proficient in assigning appropriate repertoire, tempo, and 
practice plans, not every learner has access to a qualified teacher.
Hence, there is cause to try and model the difficulty of a piece and its parts automatically.

Just as there are under-served students, there are under-served instruments.
The topic of difficulty estimation, together with fingering estimation, has been explored on the piano \cite{NAKAMURA202068, srivatsan2022checklistmodelsimprovedoutput} and to some extent also on string instruments \cite{StringInstrumentSegmentation, OptimumPathParadigm}; not, however, for wind instruments. 

In this paper, we propose a model of difficulty for the tenor saxophone.
We take the ``optimum path'' paradigm \cite{OptimumPathParadigm}, 
where control of musical instruments is fundamentally structured into ``play states'' 
or ``play actions'' meant to sound the desired tone at the desired time.
Playing a part can then be modelled as traversing a path through the corresponding play states in time,
and difficulty of a part is the cost of this traversal.

In order for this model to generalise to unseen parts, 
we can factorize the cost of a path to the aggregated costs of sub-sequences 
for which we have cost estimates, similarly to how n-gram language models truncate history. 
These sub-sequences can be as short as individual notes (unigrams); 
however, playing one note with no context is nearly meaningless. Together with \cite{OptimumPathParadigm} we settle on transitions between two fingerings as the units for which cost is defined.\footnote{There is only a small number of scenarios on a saxophone where a performer would alter fingerings based on anything other than the preceding and following fingerings.}
We retain at least some meaningful musical context, 
while keeping the state space manageable (approx.~750 possible fingering pairs on the tenor saxophone).
Traversal through a woodwind part is then the traversal through a finite-state automaton with tones as nodes and transition costs on the edges, which is easily tractable.

Two things thus must be done: defining the play states for the saxophone, 
and estimating transition costs.
For the saxophone, a play state that produces a desired tone 
is a combination of fingering, breath, muscle voicing, and articulation.
Breath, however, is hard to observe and formalise, and in any case is not entirely independent from fingering, though some characterisation of breath parameters from woodwind audio has been done \cite{kuroda2022sensing,munoz2016estimation}.
Articulation and muscle voicing for saxophones have been studied \cite{SaxophoneMuscles,scavone2008measurement}, but not to a level that they can be easily modelled computationally. Hence, we focus on encoding play states via fingering.

The final missing piece of the puzzle is then to estimate the costs of transitions.
This is, however, non-trivial for the saxophone. 
For instance, the major second from (written) C to D on a tenor saxophone varies wildly in breath support, voicing requirements, and motor movements in the three octaves where this interval appears.
Thus, in order to estimate the transition costs, a combination
of expert knowledge and real-world data is necessary.
Direct data-driven difficulty estimation \cite{DifficultyEstimation} trains a model on extrinsic annotations of difficulty, instead of intrinsic properties of the notes that make up the performance, requiring more data with difficulty ratings.
Instead, we proposed for creating fingering and difficulty prediction models that rely on data that are easier to acquire, using trill speed as a proxy for fingering transition difficulty, taking inspiration from \cite{TrillSpeedProxy}, who evaluated theoretical fingering designs for digital instruments.

Besides designing and implementing the application of the optimum path difficulty model for the saxophone using trill speeds, 
which is general enough to apply to any wind instrument without having to adjust important assumptions (such as single-voice parts\footnote{Multiphonics would rarely apply to likely users of such a system.}), 
we also contribute a dataset of tenor saxophone trills and extracted trill speeds for all its playable transitions, 
we show that expert knowledge in representation design for the inherently limited data leads to more accurate transition cost estimation as well as for minimising the data acquisition costs for more instrument models,
and we provide the implementation of the difficulty model itself as well as the pipeline for data collection and trill speed extraction.\footnote{See Section \ref{sec:data_access} for link to implementation.}

\section{Related work}
\label{sec:relatedwork}

Existing music difficulty estimation methods rely on using musician-made annotations for optimal fingerings \cite{NAKAMURA202068}, or setting (or learning) of weights based on expert observations \cite{OptimumPathParadigm}. 
For woodwinds and brass instruments, however, where large leaps can be accomplished with minimal motor movement in the fingers and arms, the rules-based approach of \cite{OptimumPathParadigm} make creating expert rules that determine optimal fingerings difficult.

Data-driven difficulty estimation methods for the piano are available, 
based on symbolic representations \cite{sebastien2012score,chiu2012study,ghatas2022hybrid},
sheet music images \cite{ramoneda2023predicting},
or recently audio \cite{ramoneda2025can};
re-arrangement by difficulty has also been done \cite{suzuki2023piano}.
For the violin, optimal fingering estimations exist: the original optimal path approach
\cite{OptimumPathParadigm}, though focus has been more on player skill level rather than part difficulty, both for fingering estimation \cite{nagata2014violin} and in context of visual perception of violinists \cite{damato2020understanding}.
In \cite{deconto2023score}, symbolic score difficulty was estimated for violin, piano, and guitar.

For the woodwind instrument family, however, 
focus has been more on their acoustical properties
\cite{wolfe2001acoustic,kuroda2022sensing}.
Saxophone acoustics have been modeled \cite{Smyth2013}
and reed parameters have been estimated from audio \cite{munoz2016estimation}.
The kinetics of fingerings and note transitions were studied for the flute \cite{almeida2009kinetics}.
In \cite{han2014hierarchical}, recorder audio is classified by typical mistake, but not difficulty. Sight-reading skills have been described for the clarinet (and violin) already in 1953 \cite{thomson1953analysis}, and for the flute they have been empirically characterised \cite{thompson1987music}, together with the difference in breath control between beginners and experts \cite{de2008analysis}.

\section{Data collection}
\label{sec:datacollection}

The process of collecting data consists of recording tenor saxophone (henceforth just saxophone) players trilling.
The musicians are instructed to trill (1) as fast as possible, (2) as cleanly as possible,\footnote{Additional clarification was given to performers to play in such a way to minimise over/underblowing that would result in hearing notes in the harmonic series of the fundamental note being played.} and (3) as \textit{legato} as possible.\footnote{On a saxophone, this would mean that the tongue is not used to separate the two notes. For larger jumps such as an octave, or when needing to play a low note that is easier to produce when tongued, this instruction was de-emphasised.}

\begin{figure}
    \centering
    \includegraphics[alt={Example trill prompt, showing a staff of sheet music with two notes in the order A, B, A. The notes have their specific fingering name written near them, with a tremolo symbol between the first two notes. Additional text is present to uniquely identify the session and trill number.},width=1\linewidth]{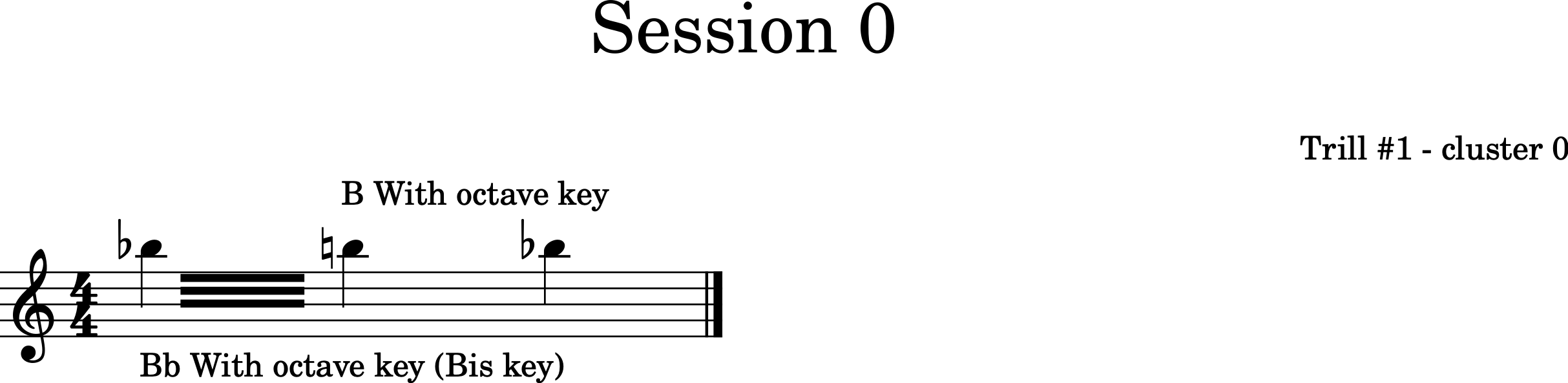}
    \caption{An example of a prompt presented to performers. In this case, the performer is meant to trill between Bb4 and B4 (written pitch).}
    \label{fig:ExamplePrompt}
\end{figure}

The data was recorded in sessions of approx. 65 trills, 
taking an hour to record (including a short break).
For each session, a set of music notation prompts -- see Figure~  \ref{fig:ExamplePrompt} -- was typeset for the players to play from, communicating which notes were to be trilled between and which fingerings were to be used for each, along with emphasising the idea of speed.

\begin{figure}[t]
    \centering
    \includegraphics[alt={Graph showing the different trill speeds for each ``anchor'' interval that was recorded during every session. Every transition has around 2 trills per second difference between the slowest and fastest recording. The slower transitions have less variance in their min and max speeds.},width=1\linewidth]{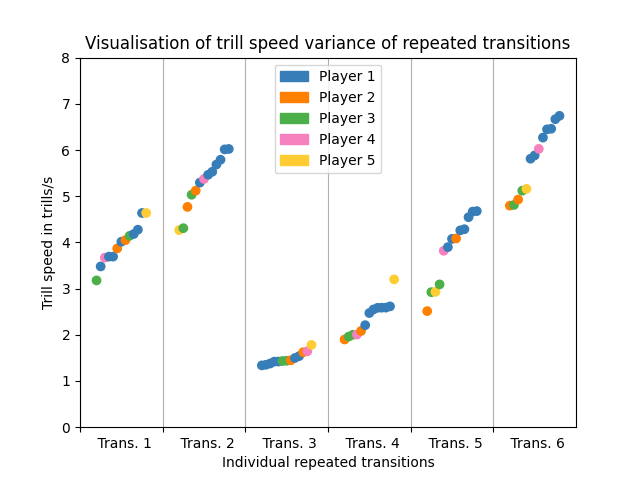}
    \caption{Visualisation of the per-player/session differences in trill speed for transitions that were recorded in every session (``anchor'' transitions).}
    \label{fig:varianceofanchors}
\end{figure}

To avoid player fatigue in a recording session, 
a very simplified prior estimate of difficulty was done using only the size of the jump and primitive information about how many fingers move, to mix likely easy and difficult trills.

Each session was subsequently manually split into recordings
of individual trills.

From the fingerings chosen, there are 741 possible fingering pairs.\footnote{Some of these were recorded but not used in future sections, as they were trills between alternate fingerings of the same note. The trill extraction methodology in Section \ref{sec:trillextraction} cannot work with this for now.}
In total, \textbf{817 trills} were recorded, with 5 different, similarly advanced conservatory-level performers participating for a total of 13 sessions.\footnote{Alternate fingerings for F\#, Bb with/without octave key, and Front F/F\# were used. These were settled on following discussion with the participating saxophonists as to which alternate fingerings they used regularly.}

6 ``anchor'' transitions were recorded in every session to explore the nature of trill speed variations across performers and sessions. The speeds for these transitions is shown in Figure~\ref{fig:varianceofanchors}. Note that Player 5 is among the fastest in 3, and among the slowest in the other 3.

\begin{figure}
    \centering
    \includegraphics[alt={A spectrogram-like chart showing only f0 prediction. There are two seperate notes clearly visible with some slight noise in the transition period between them. The process of splitting f0 predictions into individual trills is shown.},width=1\linewidth]{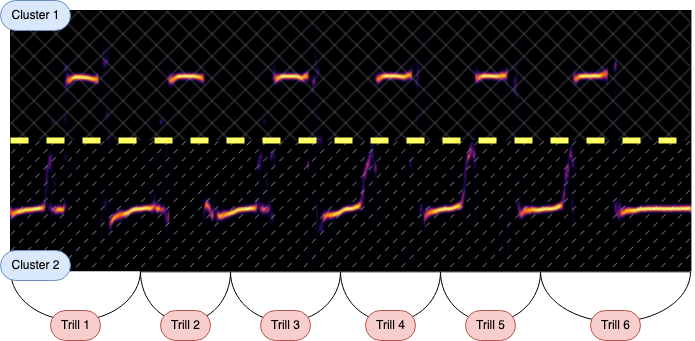}
    \caption{Example of how clusters and trills are detected visualised on a CREPE \cite{CREPE} salience plot. Dashed yellow line loosely represents the cluster boundary.}
    \label{fig:TrillAnalysis}
\end{figure}

\section{Trill Speed Extraction Workflow}
\label{sec:trillextraction}

After recording all necessary trills,
an automatic trill speed extraction step is run to extract the highest trill speed from the raw audio for each fingering transition.

We use CREPE \cite{CREPE} to predict $f_0$ from the raw audio. To determine what notes participate in the trill, we perform k-means clustering on the $f_0$ values with $k=2$. The MIDI values are then derived from the median value of each cluster using \texttt{librosa} \cite{Librosa0.10.2.post1}, at A=440 Hz.

The trill speed for each segment was then determined by the number of complete trills completed within that segment (see Figure \ref{fig:TrillAnalysis}). A complete trill is defined as starting on note 1, transitioning to note 2,
and returning to note 1.
In practice, performers vary speed over each trill. 
To find the fastest stable trill speed, we take the first half, middle two quarters, and second half of each trill (each segment is about 1.5--3 seconds long -- sufficient to count as a sustained trill), and take the highest average trill speed from these three segments.

On more technically difficult transitions at speed, under and overblowing of the note happens.
To correct for possible misclusterings, where such mistakes 
are detected as their own cluster,\footnote{Especially on sub-fifth transitions, where an octave distance from the true notes might ``tempt'' k-means to put both target notes in one cluster.} 
if one cluster was less than $20 \%$ the size of the other,
we discarded it and re-ran k-means with $k=2$ again.

To check for possible errors in trill speed extraction, the detected notes are matched against the expected interval from the recording session.
If the notes detected by the trill speed extraction algorithm do not match the expected notes,
we checked the trill recording manually to make sure the extracted trill speed does match the recorded trill.

\section{Transition features}
\label{sec:features}

\begin{figure}
    \centering
    \includegraphics[alt={Saxophone fingering chart with an adjacent example of a fingering encoding. Lines are drawn between the encoding and the fingering chart to explain the relationship between encoding and key. Keys on the fingering chart have numbers to show which position in the encoding they correspond to.},width=0.6\linewidth]{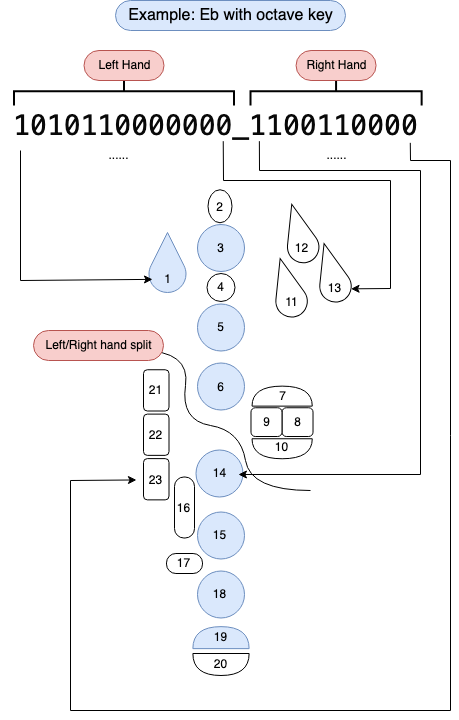}
    \caption{A diagram explaining the encoding format as it relates to a fingering chart. 1 (octave key) -- 13 are left hand keys, 14--23 are right hand keys.}
    \label{fig:fingeringencoding}
\end{figure}

To estimate costs for unrecorded transitions,\footnote{Which occur if we subsampled as per Section \ref{sec:sampling}.} a representation of transitions is needed.
Each transition is a pair of play states; we represent play states
as fingerings. The saxophone has a finite number of keys (23, see Figure~\ref{fig:fingeringencoding}) that can be pressed, so 
we can straightforwardly represent a transition as two binary fingering vectors.\footnote{For instruments where a large amount of adjustments can be made via the embouchure or other non-finger motor movements that are known for each pitch, these would also be incorporated.} However, while this representation does in principle contain all information about a transition,
more recorded data points may be needed to learn how the keys interact -- which combinations
are difficult and which are easy. 
Because we want to minimise the amount of recording necessary to reach a given cost estimation accuracy, 
we design two more representations:
one that focuses on the player's fingers rather than the instrument's keys,
and expert features that incorporate factors of difficulty from saxophone pedagogy.

\begin{figure}
    \centering
    \includegraphics[alt={Diagram showing how different feature sets are derived from a given transition. The features have annotations that explain the way they tie back to motor functions of the fingers/hands.},width=1\linewidth]{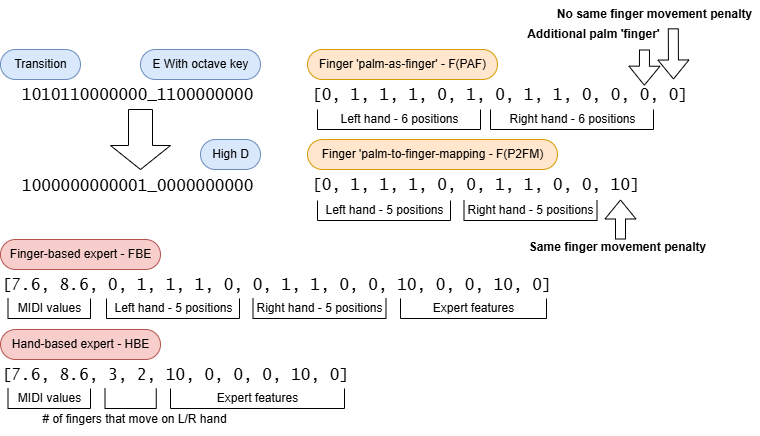}
    \caption{Encoding a transition (E with octave key to high D) with Raw,  Expert feature sets are in the version that includes MIDI and with expert weights enabled.}
    \label{fig:palm-key-approaches}
\end{figure}

In total, three representations are used for transitions (shown in Figure~\ref{fig:palm-key-approaches}):

\begin{itemize}[noitemsep]
    \item ``Raw'' (abbr. R) features -- raw encodings are simply concatenated.
    \item ``Finger'' (abbr. F) features -- each finger gets a 1 or 0 based on whether it has to move during the transition, along with a penalty if any finger has to move from one pressed key to a different pressed key (presence of a same-finger transition). 
    \item ``Expert'' features -- further divided into ``Hand-Based'' (E-HB)  and ``Finger-Based'' (E-FB).
\end{itemize}

\subsection{Finger features}
\label{sec:palm-as-a-finger}

Each combination of covered keys has a sufficiently unique map to which fingers are active,
so the ``Finger'' features can be derived automatically from the ``Raw'' features.

The only complication is dealing with palm keys (played by the side of the hand or fingers). 
Two methods of mapping keys to fingers were used for the Finger feature set: a 'palm-as-a-finger' (PAF) approach, where palm keys are treated as being played by a 'sixth finger' on each hand representing the palm, and a 'palm-key-to-finger-mapping' (P2FM) approach, where each palm key is mapped onto the finger that plays it using its side. For example, the High D key (key 13 in Figure \ref{fig:fingeringencoding}) is played by the side of the index finger.

\subsection{Expert features}
\label{subsec:expert-features}

The ``Expert'' features, selected in consultation with the recording saxophonists, contain:

\begin{itemize}[noitemsep]
    \item MIDI values of both fingerings (written pitch).
    \item Number of fingers that have to move from a pressed key to a different pressed position (so called same-finger transitions). These are big indicators of difficulty, as they are impossible to play legato and are generally cumbersome.\footnote{The Bis key is ignored for this, as it is a special case of an intended same finger transition.}
    \item Presence of any palm key transition on the left and right hand respectively.
    \item Whether or not the octave key has to change state.
    \item Presence of a fingering that is below a low C\#. These notes require additional embouchure and air support.
\end{itemize}

To this shared base, the Hand-Based Expert feature set (E-HB) adds the number of fingers that have to change on the left and right hand respectively. A transition requiring the left pointer, middle and ring finger to change states would receive a value of 3 for this feature (for the left hand).
The Finger-Based Expert features (E-FB) instead add a feature for every finger, with value 1 if it has to move and 0 otherwise.\footnote{This part is almost identical to the finger-based features.}
We experimented also with omitting the MIDI values when training a model (labelled ``NoM''). 

In addition, Expert Weights (EW) were chosen to help indicate how severely a given feature is likely to impact the trill speed for a given transition. These weights were manually adjusted until k-means clustering with $k=n/5$\footnote{Chosen as an amount of transitions comparable intuitively.} resulted in clusters of approximately the same difficulty as judged by the expert. We experimented with disabling these as well (labelled ``NoEW'').

\begin{table*}[!htb]
    \centering
    \begin{tabular}{|c|c|c|c|c|c|c|}
        \hline
        \textbf{Feature set} & \small \textbf{LM}(MSE) & \small \textbf{LM}(wMSE) & \small \textbf{LM}(MAPE) & \small \textbf{MLP}(MSE) & \small \textbf{MLP}(wMSE) & \small \textbf{MLP}(MAPE)\\
        \hline
        R & 1.06 & 1.95 & 0.39 & 0.65 & 1.41 & 0.25\\
        \hline
        F(PAF) & 0.97 & 2.02 & 0.39 & 0.52 & 0.96 & 0.25\\
        \hline
        F(P2FM) & 0.98 & 2.10 & 0.39 & 0.53 & 1.06 & 0.25\\
        \hline
        E-HB & 0.94 & 1.93 & 0.39 & 0.42 & \textbf{\textit{0.56}} & 0.22\\
        \hline
        E-HB(NoM) & 0.94 & 1.91 & 0.39 & \textbf{\textit{0.34}} & \textbf{0.53} & 0.20\\
        \hline
        E-HB(NoEW) & 0.94 & 1.94 & 0.39 & \textbf{\textit{0.36}} & 0.79 & \textbf{0.18}\\
        \hline
        E-HB(NoM\&EW) & 0.94 & 1.91 & 0.39 & \textbf{\textit{0.35}} & 0.67 & \textbf{\textit{0.19}}\\
        \hline
        E-FB & 0.94 & 1.92 & 0.38 & 0.63 & 1.16 & 0.26\\
        \hline
        E-FB(NoM) & 0.94 & 1.90 & 0.39 & 0.47 & 0.90 & 0.22\\
        \hline 
        E-FB(NoEW) & 0.94 & 1.93 & 0.38 & 0.45 & 1.23 & \textbf{\textit{0.19}}\\
        \hline
        E-FB(NoM\&EW) & 0.94 & 1.90 & 0.39 & \textbf{0.33} & 0.75 & \textbf{0.18}\\
        \hline
    \end{tabular}
    \caption{Table showing the average MSE, average WJD-weighted MSE (wMSE), and average MAPE over all folds for a given model and feature extraction method (Best per column in bold; runner-ups in bold italics). Note that MAPE of e.g. 0.2 corresponds to the estimate being 20\% off on average.}
    \label{tab:lmvsmlp}
\end{table*}

\section{Transition Difficulty Estimation}
\label{sec:modelcomparison}

To estimate the difficulty of transitions (intervals) that are not recorded, we train regression models. This step can significantly lower the cost of parameterising the entire difficulty model, because it can decrease the number of recording sessions needed.
There are thus two goals: \\(1) to obtain a model that is as accurate in predicting transition trill speeds as possible, with (2) as few of the recorded transitions as possible used as training data.

Experiments were conducted to compare the proposed feature sets and two model classes: a linear regression model (LM), and a multi-layer perceptron (MLP), both using \texttt{sklearn} \cite{sklearn}. 
The MLP had a hidden layer size of 50, and used the \texttt{lbfgs} solver. 
Both of these models additionally clamped any prediction below 0.5 trills/s to 0.5.\footnote{0.5 is still lower than any actual recorded trill speed.}

A stratified $k$-fold with shuffle was used to divide the data into folds of 150 test trills, with stratification classes dictated by binning transitions by their trill speed (bins 0--1.5, 1.5--3, 3--4.5, 4.5+). 

We report mean squared error (MSE), and also weighted mean squared error (wMSE) weighted by the relative frequency of intervals in the Weimar Jazz Database (WJD) \cite{Weimar}, to measure how the model performs with respect to what one in fact encounters in repertoire rather than with respect to what the instrument can play (although we are more interested in the latter).

Besides mean squared error, we also evaluate using mean absolute percentage error (MAPE). For the use case of difficulty estimation, a predicted and actual trill speed of 1 and 2 is a much larger issue than a similar absolute difference of 5 and 6. Additionally, the easier a transition is (and thus possessing a higher recorded trill speed), the more variance there may be between performers and even between recordings of similar transitions by the same performer, as, at the higher ends of the speed scale, fatigue and other factors play a much larger role; such a phenomenon can be seen in Figure~\ref{fig:varianceofanchors}.

Results of trill speed estimation experiments are in Table~\ref{tab:lmvsmlp}.
As evidenced by the mean squared error (MSE) values found during the experiment, the LM stops improving at $\pm0.97$ MSE and 0.39 MAPE.
The MLP performed much better, getting down to $\pm 0.4$ MSE and just under 0.2 MAPE using the Expert feature sets.
The Finger-based feature set performed better than the Raw feature set
in MSE but not in MAPE, suggesting that the improvement in MSE comes in faster trills -- which is congruent with these features more closely related to motor limitations.

\begin{figure}
    \centering
    \includegraphics[alt={Graph that shows the relationship between true and predicted trill speed for the Expert Hand-based model. The predicted trill speeds are accurate for the slower transitions, but there is an increase in mistakes as the transitions get faster.},width=1\linewidth]{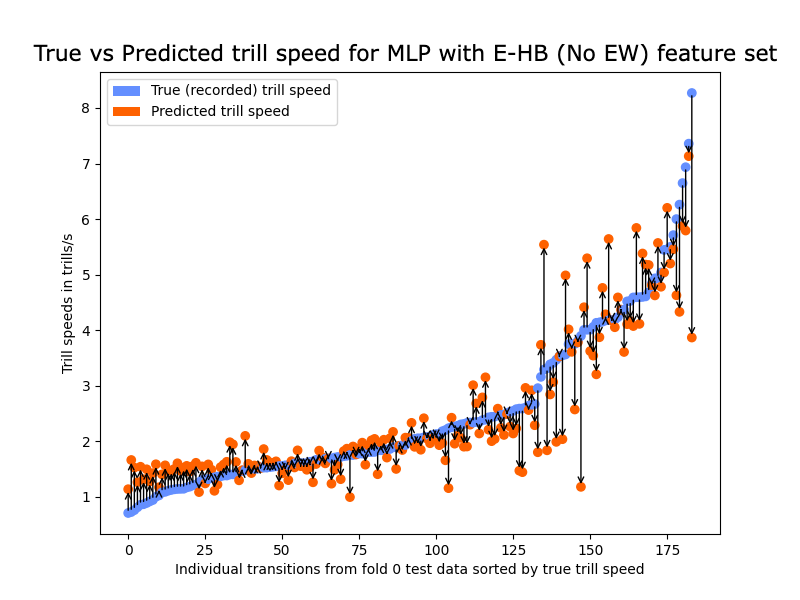}
    \caption{Differences between true and predicted trill speeds for the MLP model using the E-HB(NoEW) feature set. Transitions are sorted in ascending order by their true trill speed.}
    \label{fig:bestModelErrorGraph}
\end{figure}

Errors in wMSE are larger than equally weighted MSE;
easier (faster) transitions are more common, so the more heavily weighed errors are on transitions with more noise.

The main takeaway, however, is that expert-designed features clearly perform best,
probably as a function of the small maximum available dataset size.
At the same time, however, the relationships between individual features are non-trivial 
in any representation, as the MLP clearly performs better than the linear model.
We estimate a MAPE lower bound from the ``anchor'' transitions (Figure \ref{fig:varianceofanchors}), with the median speed of each transition as its the `true' value. The average anchor MAPE was 0.10 (max. 0.14, min. 0.06). The best model is thus only 8\% worse than the attainable maximum.

\section{Sampling methods}
\label{sec:sampling}

Not all recorded transitions are equally informative, so a good sampling strategy might reduce the number of recordings necessary to achieve a target prediction error.

For the following tests, the Expert Hand-Based (No Expert Weights) features were used.\footnote{Chosen as it is in the slightly better performant expert feature set in the form that achieved the best MAPE.}
For each sampling method, the stratified k-fold cross-validation from Figure~\ref{sec:modelcomparison}, was used. For every fold, the training set was then downsampled to the target $n$ 
using the given sampling method. 
On this downsampled training set the MLP from Figure~\ref{sec:modelcomparison} was trained, 
and evaluated on the test set. For each $n$, 3 independent, differently seeded samples were tested, 
with the graphed value being the mean of the MAPEs of each sampling attempt.
We compared three sampling methods:

\textbf{Uniform} sampling of recorded transitions.

\textbf{Cluster-based.} For every data point, its Hand-based Expert features (with Expert weights\footnote{The expert weights were specifically designed in theory for this clustering, not for the model.} and MIDI enabled) were extracted. Then, a $k$-means clustering was run, with $k=n$, so that we would get one cluster per sample point we want to get, and from each cluster a random data point was uniformly selected into the training sample.

\textbf{Empirical.} We extracted note bigram probabilities from the Weimar Jazz Database \cite{Weimar} (WJD),
with Laplace smoothing at $\alpha = 0.1$,
and sampled the transitions according to these probabilities.
For notes with multiple fingerings, possible transitions were sampled uniformly.
We expect this sampling to perform worse -- but it roughly speaks to the data efficiency of just recording players play repertoire and extracting transition costs from that.

\begin{figure}
    \centering
    \includegraphics[alt={Figure shows three different coloured lines, each representing a sampling method. Empirical sampling is by far the worst, at around 0.5 MAPE at 150 samples, though it does converges to the same flat line as the number of samples goes closer to 500. Cluster sampling and uniform sampling are similar, with cluster being slightly better at lower n.},width=1\linewidth]{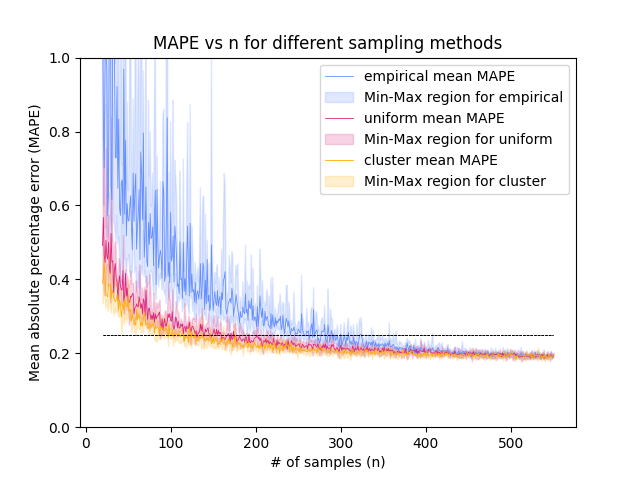}
    \caption{Graph showing MAPE as training sample size increases across sampling methods. The coloured lines are the mean MAPEs over all folds at a given sample size.  
    0.25 mean MAPE is shown as a possible target MAPE choice at which to compare sampling strategies.}
    \label{fig:samplingMAPE}
\end{figure}

As seen in Figure~\ref{fig:samplingMAPE}, cluster-based sampling performs best, with uniform sampling coming in a close second. Cluster-based sampling is slightly better at lower $n$'s, as it likely creates a more informative set of training data. However, the differences are in practice minimal.
Empirical sampling performs much worse; frequently occurring transitions are not more informative.

While the uniform and cluster sampling never diverge too much, the shallow slope somewhat means that around the region of MAPE 0.25, the difference between these two sampling methods is about one recording session worth of recordings -- 2 for clusters vs. 3 for uniform. Again, expert features show their usefulness.

\section{Application Example}
\label{sec:applications}

Estimating the difficulty of a saxophone part 
is finally done by finding the optimal path through a finite-state automaton 
between our first and last set of states. 
The edge weights are the maximum attainable trill speeds estimated by the MLP model with E-HB(NoEW) features.
We do not use the measured trill speeds directly because we want our system to also be usable for other instruments for which the estimation methodology described above allowed proceeding without recording all the transitions.

Given that note values are known from the part, we can account for a given target tempo. 
The ``fastest'' (maximal) path through the part in terms of fingerings is found with Viterbi decoding over the estimated trill speeds (yielding optimal fingerings). Then, for every fingering bigram in the part, the transition difficulty
is the proportion of the required ``half trill'' speed at the target tempo to the estimated maximum trill speed for that bigram.

\begin{figure}
    \centering
    \includegraphics[alt={Sheet music of a short musical passage for tenor saxophone. Note heads are coloured different shades of black to red. Red is meant to represent a more difficult transition. Notable red transitions are the transition from D flat without an octave key to Eb with an octave key at speed, and low D flat to low B flat to low B natural.},width=1\linewidth, trim=2cm 23.5cm 1cm 4cm, clip]{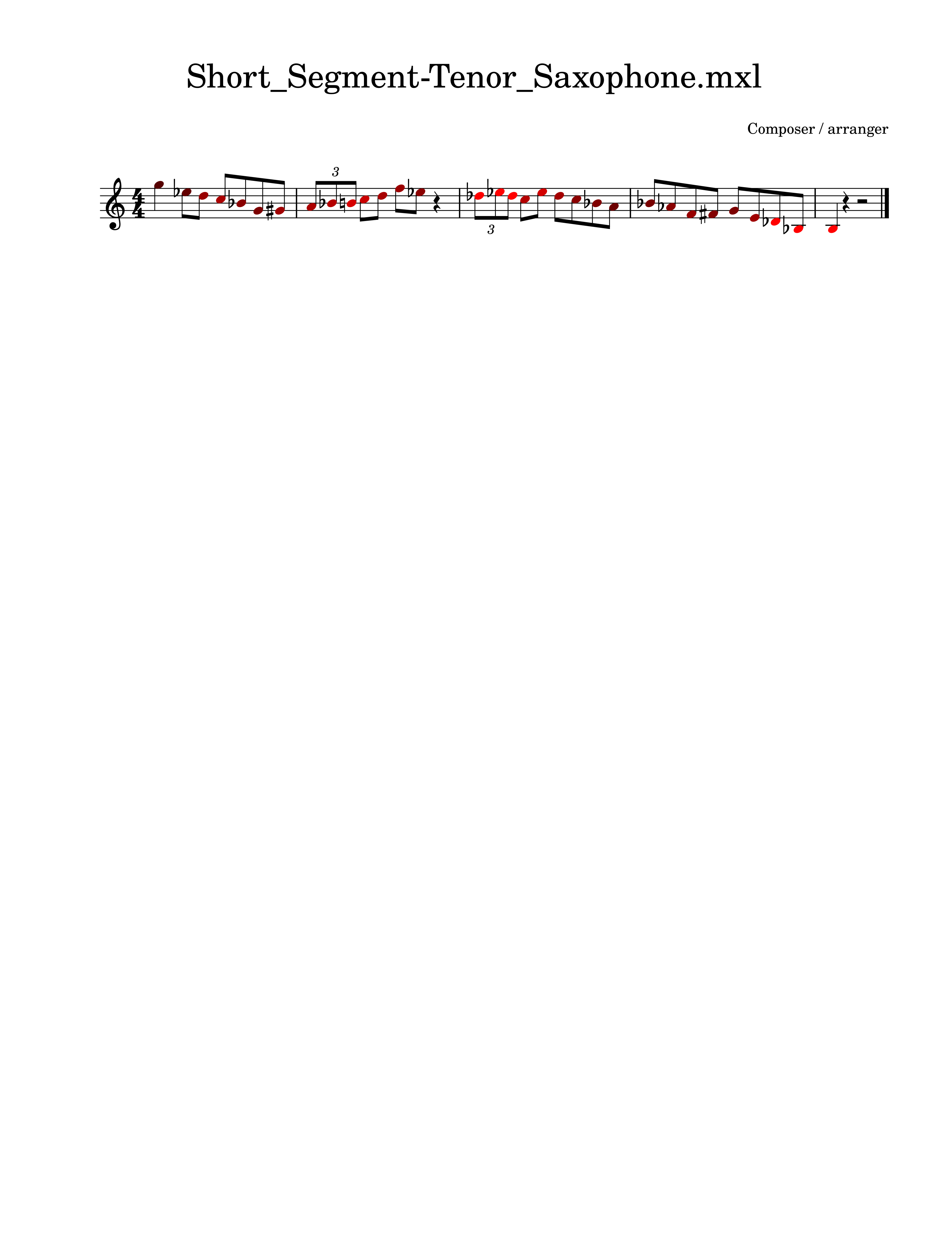}
    \caption{Example difficulty visualisation tool output. The redder a note is, the more difficult the segment around is deemed to be. Tempo for this visualisation was set to 160 BPM, which is not particularly fast for jazz \cite{Collier1994}.}
    \label{fig:difficulty-visualiser}
\end{figure}

The outcome of one such model pass is visualised in Figure~\ref{fig:difficulty-visualiser}.\footnote{Two full \texttt{*.mxl} examples are available in the Github repository (see Section \ref{sec:data_access}).}
The brightness of each notehead is given by the average of its incoming and outgoing transition difficulties.
This excerpt shows the non-trivial difficulty structure of the saxophone.
The low notes are harder than mid-range especially due to complex pinky finger movement at the end of the phrase. At the same time it is sensitive to a major factor of technical difficulty -- speed, with the triplets standing out.
We created a Musescore 3.6 plugin implementing this visualisation (see Section \ref{sec:data_access}).

\section{Discussion \& Conclusion}

We have proposed a method for estimating the difficulty of woodwind parts based on the ``optimal path'' paradigm \cite{OptimumPathParadigm} 
and implemented it for the tenor saxophone, providing data, methods for transition weight estimation, and showing how expert knowledge can help minimise the cost of creating such models for other instruments. It is an example of participatory, practice-based research.

However, there are also significant limitations.
Most importantly, we would prefer to empirically evaluate the overall estimation, but we lack digitally encoded saxophone parts with authoritative ground truth difficulty. The ideal source would be piece grading from a musical examination board such as the ABRSM (Associated Board of the Royal Schools of Music), but digital encodings of the ABRSM-graded pieces are not available and copyright issues would prevent sharing an evaluation corpus, so difficulty evaluation data must be gathered.\footnote{This also speaks to the under-representation of woodwinds in MIR; although an audio-to-score system for the saxophone has been tried \cite{MartnezSevilla2023}.}

Trill speed
does not directly capture the intricacies of voicing, articulation, and breath support, which are all major contributors to woodwind difficulty. To maintain discrete play states, knowledge of these mechanics could be encoded in additional expert features. 
Also, trill speeds can be asymmetrical. On the saxophone, going up an octave quickly is much easier than down an octave;
this is not modeled.

This methodology still requires initial access to a technically proficient instrumentalist (and domain expert) to do the recordings and design the expert features and instrument encodings. Especially when using cluster-based sampling, many of the recorded transitions may be large, technical leaps that may be difficult for beginners to play.

As can be seen in Figure~\ref{fig:varianceofanchors}, even for the same transition, the variance of the recorded trill speed can be high. Some normalisation procedure is needed. However,
as seen in Figure~\ref{fig:varianceofanchors}, a performer may be the fastest for one transition, but slowest in another, 
so a single coefficient per player is unlikely to be effective.
However, the opportunity here is to at the same time arrive at a normalisation function that can easily \textit{personalise} the transition costs to a specific player based solely on recording a few anchor transitions.

Despite these limitations, we believe this work can serve at least as a first step towards including the woodwind family more into MIR for music pedagogy, and we look forward to how others might take inspiration and devote some more attention to these instruments.

\section{Acknowledgments}

This work is supported by project ``Human-centred AI for a Sustainable and Adaptive Society'' (reg. no.: CZ.02.01.01/00/23\_025/0008691), co-funded by the European Union. Computing infrastructure was provided by the LINDAT/CLARIAH-CZ Research Infrastructure (https://lindat.cz), supported by the Ministry of Education, Youth and Sports of the Czech Republic (Project No. LM2023062).

\section{Ethics statement}

All players signed an informed consent form. All data is pseudonymised. Video recordings of sessions, taken as a backup source for checking against unexpected errors in extraction, were deleted as soon as the trill and trill speed extraction steps were finished and checked.
No generative AI tools were used in producing this text; code generation was used for hints in visualisations that are used as figures.

\section{Data and code accessibility}
\label{sec:data_access}

The raw audio data in the form of recorded trills is here: \url{http://hdl.handle.net/11234/1-5942}. This also includes the output of the trill speed extraction pipeline, as rerunning the pipeline for all the data is time-consuming.
All code and other implementation details can be found at \url{www.github.com/Vobludalib/SaxophoneDifficultyModel/tree/ISMIR2025}. 
The Musescore 3.6 plugin implementing this model can be found at \url{www.github.com/Vobludalib/SaxophoneDifficultyModel/tree/main/plugin} .

\end{document}